\documentclass{INTERSPEECH2023}

\usepackage{amsmath,graphicx,multirow}
\usepackage{bbding}
\usepackage{booktabs}
\usepackage{cite}
\usepackage{verbatim}
\usepackage{bibspacing}

\usepackage[utf8]{inputenc}
\usepackage[titletoc,title]{appendix}
\usepackage{makecell}
\usepackage{tipx}

\newcommand{\xmark}{\ding{55}}
\newcommand{\cmark}{\ding{51}}

\usepackage{color}
\usepackage{multirow}
\usepackage{hhline}
\usepackage{subfigure}
\usepackage{pifont}
\usepackage{bm}
\usepackage{amsmath,amsfonts,amssymb,mathtools}

\usepackage{graphicx,float}

\interspeechcameraready

\title{Exploiting Cross-Domain And Cross-Lingual Ultrasound Tongue Imaging Features For Elderly And Dysarthric Speech Recognition}
\name{Shujie Hu$^1$, Xurong Xie$^2$, Mengzhe Geng$^1$, Mingyu Cui$^1$, Jiajun Deng$^1$, Guinan Li$^1$, Tianzi Wang$^1$, Helen Meng$^1$, Xunying Liu$^1$}
\address{
  $^1$The Chinese University of Hong Kong, Hong Kong SAR, China\\
  $^2$Institute of Software, Chinese Academy of Sciences, China}
\email{\{sjhu,mzgeng,mycui,jjdeng,gnli,twang,hmmeng,xyliu\}@se.cuhk.edu.hk, xurong@iscas.ac.cn}

\begin{document}
\bstctlcite{IEEEexample:BSTcontrol}
\maketitle

\begin{abstract}
Articulatory features (AFs) are inherently invariant to acoustic signal distortion. Their practical application to atypical domains such as elderly, disordered speech across languages is limited by data scarcity. This paper presents a cross-domain and cross-lingual Acoustic-to-Articulatory (A2A) inversion approach that utilizes the parallel audio and ultrasound tongue imaging (UTI) data of the 24-hour TaL corpus in A2A model training before being adapted to three datasets: the English DementiaBank Pitt and Cantonese JCCOCC MoCA elderly speech corpora; and the English TORGO dysarthric speech data, to produce UTI based AFs. Experiments suggest incorporating the generated AFs consistently outperforms the baseline TDNN/Conformer ASR systems using acoustic features only by statistically significant word/character error rate reductions up to 4.75\%, 2.59\% and 2.07\% absolute (14.69\%, 10.64\% and 22.72\% relative) after data augmentation, speaker adaptation and cross system multi-pass decoding.
\end{abstract}

\vspace{-0.2cm}
\section{Introduction}
\vspace{-0.1cm}
\begin{figure*}[htbp]
  \centering
  \begin{minipage}[t]{0.44\linewidth}
    \centering
    \includegraphics[width=\linewidth]{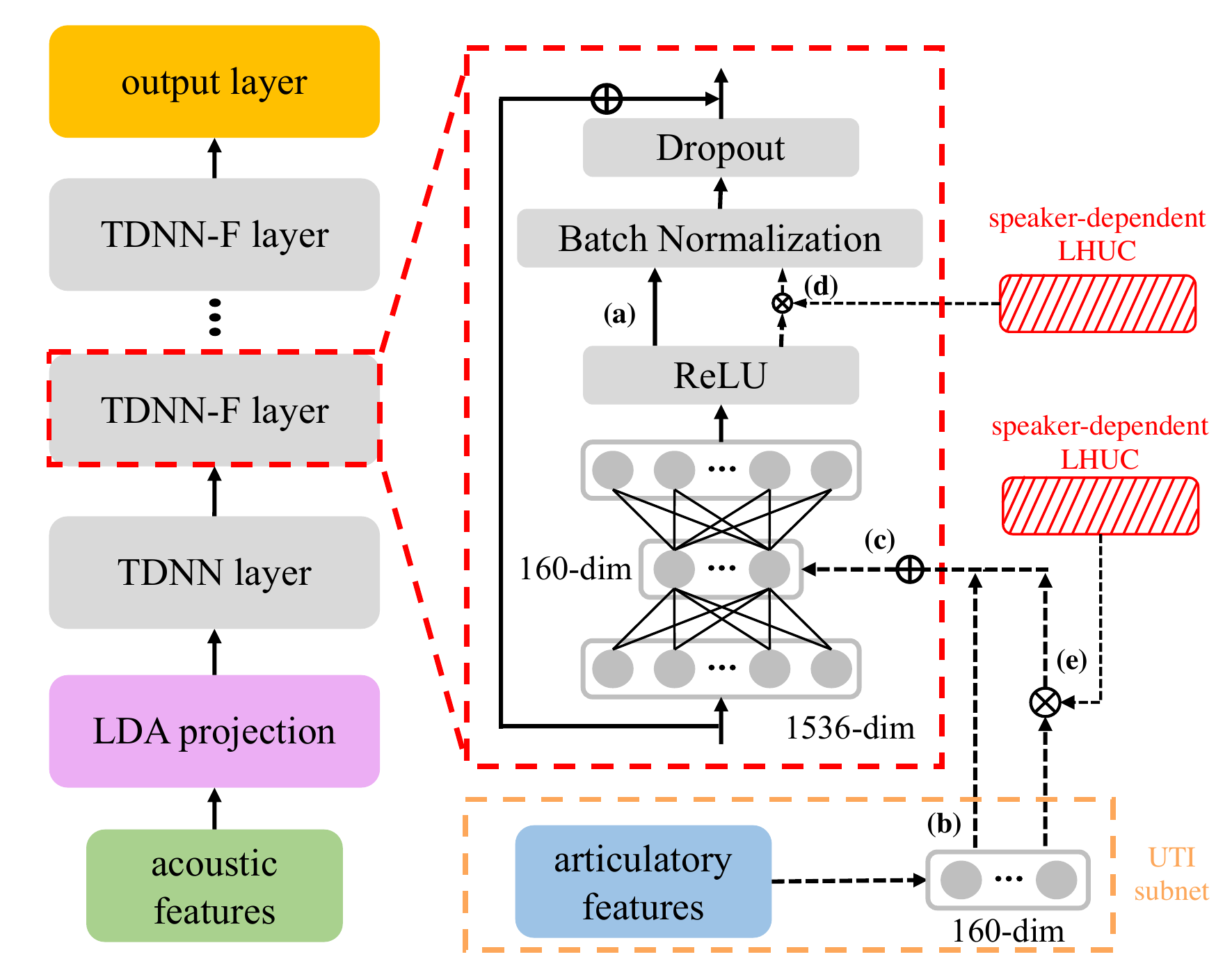}
    \label{fig:side:a}
  \end{minipage}%
  \begin{minipage}[t]{0.44\linewidth}
    \centering
    \includegraphics[width=\linewidth]{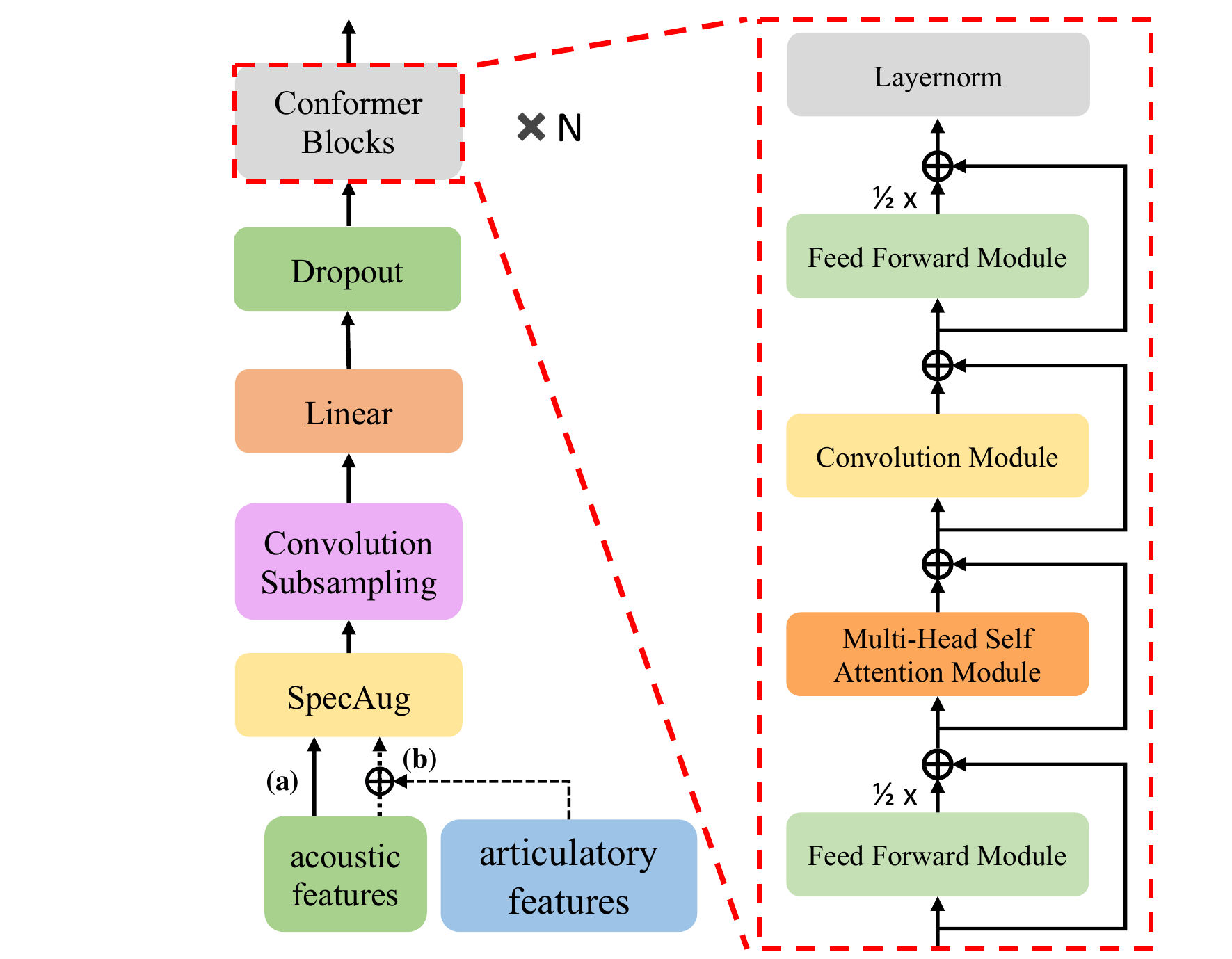}
    \label{fig:side:b}
  \end{minipage}
  \vspace{-0.5cm}
  \caption{The left part of the Figure is the factored TDNN based ASR and AASR system architecture. Different connection combinations in this figure form different systems. Retaining connection (a) while discarding others leads to the ASR baseline system; Retaining (a), (b) and (c)(for articulatory modality fusion) produces the TDNN AASR system. The connections (d) and (d)+(e) are used for LHUC speaker adaptive training of ASR and AASR systems. The right part of the Figure shows the Conformer based end-to-end system. The AASR system uses the acoustic-articulatory concatenation via connection (b).}
  \vspace{-0.6cm}
\end{figure*}
Despite the rapid progress of automatic speech recognition (ASR) technologies in the past few decades, accurate recognition of elderly and disordered speech remains a challenging task \cite{yu2018development,xiong2020source,liu2021recent,ye2021development,geng2022speaker, yue2022acoustic, hu2022exploring}. 
Neurocognitive disorders, such as Alzheimer’s disease (AD), are often found among older adults and manifest themselves in speech and language impairments including weakened neuro-motor control in speech production and imprecise articulation. 
ASR based assistive technology developments tendering for such users’ needs play a vital role in not only improving their quality of life and social inclusion, but also facilitating large scale automatic speech based early diagnosis of neurocognitive impairment and preventive care \cite{ferri2005global,rudzicz2014speech,zhou2016speech,mirheidari2019dementia,ye2021development}.
\par
Elderly and disordered speech bring challenges on all fronts to current deep learning based ASR technologies predominantly targeting non-aged, healthy adult users. First, a large mismatch between such data and non-aged, healthy adult voice is often observed. Such difference manifests itself across many fronts including articulatory imprecision, decreased volume and clarity, changes in pitch, increased dysfluencies and slower speaking rate.
Second, the co-occurring disabilities, mobility or accessibility limitations often found among elderly and disordered speakers lead to the difficulty in collecting large quantities of such data that are essential for current data intensive ASR system development. In addition, sources of variability commonly found in normal speech including accent or gender, when further compounded with those over age and speech pathology severity, create large diversity among speakers \cite{kodrasi2020spectro,smith1987temporal}. 
\par
Human speech production involves the coordinated movements of various articulators such as the tongue, lips, teeth and palate. Articulatory movement features are inherently invariant to extrinsic acoustic distortion. They have been successfully applied to normal \cite{zlokarnik1995adding,kirchhoff2002combining,mitra2017hybrid} and pathological speech \cite{rudzicz2010articulatory, gonzalez2017direct, xiong2018deep, yilmaz2019articulatory} recognition. In practice, recording detailed articulatory movements and vocal tract shape normally requires the use of intrusive electromagnetic articulography (EMA) \cite{engwall2000static} or magnetic resonance imaging (MRI) \cite{narayanan2014real} techniques. Compared to EMA  and MRI, ultrasound tongue imaging (UTI) \cite{stone2005guide, ribeiro2019speaker,cleland2019enabling} is more portable, less non-invasive and costly. UTI uses B-mode diagnostic ultrasound  to visualize the tongue surface movements during speech production at a high frame rate \cite{ribeiro2021tal}. However, there are currently very few publicly available UTI corpora, all of which were designed for the English and of limited size~\cite{cai2011recognition, eshky2019ultrasuite, ribeiro2021tal}. 
By far the largest Tongue and Lips corpus (TaL) \cite{ribeiro2021tal} contains 24 hours of parallel ultrasound, video, and audio data collected from 81 native speakers of English. The practical and wider use of UTI based articulatory features in ASR systems for both normal and atypical speech task domains, and across languages, is hindered by the scarcity of such specialist data.
\par
An alternative and general approach to obtain articulatory information is to estimate it from the more accessible acoustic speech signals using data driven artificial neural network based acoustic-to-articulatory (A2A) inversion techniques \cite{papcun1992inferring, uria2012deep, xie2018investigation, maharana2021acoustic}. As the A2A inversion model training only requires a part of training materials containing parallel acoustic-articulatory data, the resulting inversion model can be used to produce articulatory features when only audio recordings are available. A wider and more practical application of articulatory features in ASR systems becomes possible. Prior researches on A2A inversion were conducted predominantly on normal speech task domains \cite{papcun1992inferring, uria2012deep, xie2018investigation} and most of them used EMA based articulatory features. In contrast, very limited researches were carried out on A2A inversion for UTI based articulatory features \cite{porras2019dnn}.

\par
In order to address the issues mentioned above, this paper presents a cross-domain and cross-lingual A2A inversion approach that utilizes the parallel acoustic-articulatory data of the 24-hour TaL corpus \cite{ribeiro2021tal} in A2A model training before being cross-domain and cross-lingual adapted to three datasets across two languages: the English DementiaBank Pitt \cite{becker1994natural} and the Cantonese JCCOCC MoCA \cite{xuspeaker} elderly speech corpora; and the English TORGO \cite{rudzicz2012torgo} dysarthric speech data to produce UTI articulatory features. Bi-LSTM based deep neural network A2A inversion models are used. A cross-domain adaptation network is used to reduce the acoustic mismatch between the TaL corpus \cite{ribeiro2021tal} and the target elderly and disordered speech corpora. 
\par
The main contributions are summarized below. To the best of our knowledge, this work is the first systematic study of using audio and ultrasound tongue imaging parallel recordings trained A2A inversion models for UTI articulatory features generation targeting elderly and disordered speech recognition. In contrast, prior researches on using UTI features were conducted mainly in the context of non-aged adult or child speech recognition \cite{hueber2010development,ji2018updating,ribeiro2021tal,ribeiro2019ultrasound}.
In addition, this work is also the first use of cross-lingual and cross-domain UTI based articulatory features to develop elderly and disordered speech recognition systems. On all three tasks across both the English and Cantonese languages, significant WER or CER reduction is obtained over the acoustic features only baseline hybrid TDNN \cite{peddinti2015time} and E2E Conformer \cite{gulati2020conformer} systems after speed perturbation plus SpecAugment based data augmentation, learning hidden unit contributions (LHUC) \cite{swietojanski2016learning} speaker adaptation and cross system multi-pass decoding \cite{cui2022two} are applied. Further analysis of the RMSE metric and the phonetic discrimination brought by UTI articulatory features via T-SNE \cite{van2008visualizing} visualization are also given.

\vspace{-0.2cm}
\section{Articulatory Feature Based Speech Recognition}
\vspace{-0.1cm}
This section describes the time delay neural network (TDNN) \cite{peddinti2015time} or E2E Conformer \cite{gulati2020conformer} based ASR and acoustic-articulatory feature based speech recognition (AASR) system architecture on elderly and disordered corpora.

Both of the baseline hybrid TDNN ASR and AASR systems share the same main structure based on a 7-layer or 14-layer factorized TDNN (TDNN-F) model with a semi-orthogonal constraint and they are trained using the sequence discriminative lattice-free MMI (LF-MMI) criterion (the left part of Fig. 1). The TDNN-F hidden layers positioned after LDA projection and TDNN layer are shown in the red dotted box in Figure 1. The AASR system is produced by adding an ultrasound tongue imaging (UTI) subnet using connections (b) and (c). The output of the UTI subnet is then added to the 160-dimensional factorized hidden layer.
To model the large variability among elderly and disordered speakers, learning hidden unit contributions (LHUC) \cite{swietojanski2016learning} based speaker adaptive training (SAT) is used. Speaker-level LHUC scaling factors (in red) are applied to the main TDNN-F architecture via connection (d) and UTI component via connection (e) respectively (left part, Fig. 1). 
\par
E2E Conformer  ASR and AASR system architectures are shown in the right part of Fig. 1, where the details of Conformer blocks are shown in the red dotted box. UTI articulatory features are incorporated at the input layer by connection (b).

\vspace{-0.2cm}
\section{Acoustic-To-Articulatory Inversion}
\vspace{-0.1cm}
\subsection{In-domain A2A Inversion}
\vspace{-0.1cm}
The objective of acoustic-to-articulatory inversion is to learn the mapping between acoustic and UTI based articulatory feature representations. To this end, a suitable inversion model is required. In previous research, state-of-the-art A2A performance was achieved using long short term memory (LSTM) networks \cite{xie2018investigation,liu2015deep}. In this paper, the out-of-domain non-aged normal speech of the TaL dataset \cite{ribeiro2021tal} accompanied with parallel ultrasound data collected from 81 native speakers of English is used to construct Bi-LSTM based A2A inversion model. The efficacy of the A2A inversion model is evaluated by measuring the performance of the resulting AASR systems built using either the ground truth, original UTI features\footnote{The UTI based articulatory features are extracted following \cite{ribeiro2021tal} where the 2D Discrete Cosine Transform (DCT) with 12×12
coefficients to each resized ultrasound frame was applied.}, or those generated using A2A inversion. The UTI based articulatory features are used in both AASR system training and evaluation stages. 

\vspace{-0.3cm}
\begin{table}[H]
\centering
\caption{WER (\%) of ASR and AASR systems on the TaL test set.}
\vspace{-0.3cm}
\scalebox{0.8}[0.8]{
\begin{tabular}{c|c|c} 
\hline\hline
\multicolumn{3}{c}{WER (\%)}                                   \\ 
\hline
\multirow{2}{*}{ASR} & \multicolumn{2}{c}{AASR}                \\ 
\cline{2-3}
                     & original features & generated features  \\ 
\hline
8.19                 & 7.47              & 7.48                \\
\hline\hline
\end{tabular}
}
\vspace{-0.4cm}
\end{table}

\par
The results in Table 1 illustrate that: 1) Using either the original or inverted articulatory features produce performance improvements over the ASR system; 2) the AASR system using the inverted articulatory features has comparable performance compared to that using original articulatory features.
\vspace{-0.3cm}
\subsection{Cross-Domain And Cross-Lingual A2A Inversion}
\vspace{-0.2cm}
Due to the acoustic domain and language mismatch, a direct cross-domain and cross-lingual application of the A2A inversion model trained on a normal, non-aged acoustic-articulatory parallel data to the elderly and disordered speech is problematic, as was shown in the previous researches on cross-domain audio-visual inversion \cite{liu2021recent} and A2A inversion \cite{hu2022exploiting}. To this end, the large acoustic domain and language mismatch are minimized using multi-level adaptive networks (MLAN) \cite{bell2012transcription, liu2021recent,hu2022exploiting} before A2A inversion can be performed.
\par
The MLAN training process includes the following steps: 1) the first level DNN is trained with the data from the in-domain English or Cantonese language specific elderly or disordered speech; 2) the resulting in-domain speech trained DNN is then used to produce bottleneck features for the out-of-domain data of the English TaL dataset; 3) the second level DNN is trained using the out-of-domain TaL audio data concatenated with the bottleneck features computed from the previous step. The combined effect produced by these two cascaded component DNNs is such that, when feedforwarding the elderly or dysarthric data of a target language into the resulting MLAN network, the final bottleneck features produced at the second-level DNN component will exhibit smaller mismatch against those obtained by feedforwarding the English TaL acoustic data into the MLAN network. The resulting cross-domain and cross-lingual adapted bottleneck features are used in A2A inversion model training and articulatory feature generation for elderly or disordered audio data of English or Cantonese. Further detailed descriptions of the cross-domain MLAN model are in \cite{bell2012transcription, liu2021recent,hu2022exploiting}.

\begin{table*}[htbp]
\centering
\caption{The performance of baseline ASR and AASR systems using the cross-domain/lingual inverted UTI articulatory features (AFs) on 1) the English DementiaBank Pitt development (Dev) and evaluation (Eval) sets; 2) the Cantonese JCCOCC MoCA development (Dev) and evaluation (Eval) sets containing elderly speakers only; AND 3) the English dysarthric TORGO test set. ``INV'' and ``PAR'' refer to clinical investigator and elderly participant. ``Seve.'', ``Mod.'' and ``Mild'' are intelligibility levels of the dysarthric speakers: severe, moderate and mild. ``4.L'' denotes a larger TDNN model with comparable \#parameters as the AASR system (Sys. 5). AFs incorporated using ``hidden layer'' fusion and optionally further score fusion ``score A+AA''. ``$\rightarrow$'' stands for multi-pass decoding. $\dag$ denotes statistical significant (MAPSSWE, $\alpha$ = 0.05) differences obtained against the baseline ASR systems (Sys. 1, 4, 6, 9)}
\vspace{-0.2cm}
\scalebox{0.68}{
\begin{tabular}{c|c|c|c|c|c|cc|cc|c!{\vrule width \heavyrulewidth}c|c|c!{\vrule width \heavyrulewidth}c|c|c|c} 
\hline\hline
\multirow{4}{*}{\begin{tabular}[c]{@{}c@{}}Sys.\end{tabular}} & \multirow{4}{*}{Model}     & \multirow{4}{*}{MLAN} & \multirow{4}{*}{AA fusion} & \multirow{4}{*}{\begin{tabular}[c]{@{}c@{}}Data\\Aug.\end{tabular}} & \multirow{4}{*}{LHUC} & \multicolumn{5}{c!{\vrule width \heavyrulewidth}}{WER (\%)}                                           & \multicolumn{3}{c!{\vrule width \heavyrulewidth}}{CER (\%)}             & \multicolumn{4}{c}{WER (\%)}                                                                    \\ 
\cline{7-18}
&                            &                       &                            &                                                                     &                       & \multicolumn{5}{c!{\vrule width \heavyrulewidth}}{\textbf{DementiaBank Pitt}}                         & \multicolumn{3}{c!{\vrule width \heavyrulewidth}}{\textbf{JCCOCC MoCA}} & \multicolumn{4}{c}{\textbf{TORGO}}                                                              \\ 
\cline{7-18}
&                            &                       &                            &                                                                     &                       & \multicolumn{2}{c|}{Dev}                      & \multicolumn{2}{c|}{Eval}     & \multirow{2}{*}{Overall} & \multirow{2}{*}{Dev}  & \multirow{2}{*}{Eval} & \multirow{2}{*}{Overall}   & \multirow{2}{*}{Seve.} & \multirow{2}{*}{Mod.} & \multirow{2}{*}{Mild} & \multirow{2}{*}{Overall}  \\ 
\cline{7-10}
&                            &                       &                            &                                                                     &                       & PAR.                  & INV.                  & PAR.                  & INV.  &                       &                       &                       &                         &                        &                       &                       &                        \\ 
\hline
1                                                                  & \multirow{8}{*}{TDNN}      & \xmark                    & \xmark                         & \multirow{3}{*}{\xmark}                                                 & \multirow{3}{*}{\xmark}   & 50.71                 & 21.58                 & 39.16                 & 21.20 & 36.04                 & 30.89                 & 27.95                 & 29.41                   & 16.22                  & 10.31                 & 3.87                  & 11.62                  \\
2                                                                  &                            & \xmark                    & hidden layer               &                                                                     &                       & 51.82                 & 20.70                 & 39.87                 & 20.87 & 36.24                 & 30.88                 & 28.27                 & 29.57                   & 16.67                  & 10.00                 & 3.64                  & 11.73                  \\
3                                                                  &                            & \cmark                   & hidden layer               &                                                                     &                       & \textbf{48.83}$^\dag$ & \textbf{19.87}$^\dag$ & \textbf{37.31}$^\dag$ & 20.64 & \textbf{34.28}$^\dag$ & \textbf{28.86}$^\dag$ & \textbf{26.15}$^\dag$ & \textbf{27.50}$^\dag$   & 16.14                  & 6.12                  & 3.48                  & \textbf{10.61}$^\dag$  \\ 
\cline{1-1}\cline{3-18}
4                                                                 &                            & \xmark                    & \xmark                         & \multirow{3}{*}{\cmark}                                                & \multirow{3}{*}{\xmark}   & 47.93                 & 19.91                 & 36.66                 & 19.76 & 33.80                 & 26.87                 & 23.71                 & 25.28                   & 12.80                  & 8.78                  & 3.64                  & 9.47                   \\
4.L                                                                  &                            & \xmark                    & \xmark                         &  &  & 47.80                 & 21.31                 & 36.85                 & 20.53 & 34.37                 & 27.02                 & 24.06                 & 25.53                   & 13.74                  & 9.39                  & 2.94                  & 9.97                   \\
5                                                                 &                            & \cmark                   & hidden layer               &                                                                     &                       & \textbf{45.82}$^\dag$ & \textbf{19.21}$^\dag$ & \textbf{34.89}$^\dag$ & 18.42 & \textbf{32.35}$^\dag$ & \textbf{25.06}$^\dag$ & \textbf{22.88}$^\dag$ & \textbf{23.96}$^\dag$   & 12.80                  & 4.59                  & 2.71                  & \textbf{8.35}$^\dag$   \\ 
\cline{1-1}\cline{3-18}
6                                                                  &                            & \xmark                    & \xmark                         & \multirow{3}{*}{\cmark}                                                & \multirow{3}{*}{\cmark}  & 45.49                 & 19.26                 & 35.44                 & 18.42 & 32.33                 & 25.77                 & 22.94                 & 24.35                   & 12.52                  & 8.27                  & 3.25                  & 9.11                   \\
7                                                                  &                            & \cmark                   & hidden layer               &                                                                     &                       & \textbf{43.06}$^\dag$ & \textbf{18.31}$^\dag$ & \textbf{33.05}$^\dag$ & 18.09 & \textbf{30.57}$^\dag$ & \textbf{24.41}$^\dag$ & 22.38                 & \textbf{23.39}$^\dag$   & 12.20                  & 5.00                  & 2.63                  & \textbf{8.09}$^\dag$   \\
8                                                                  &                            & \cmark                   & score A+AA                 &                                                                     &                       & \textbf{41.97}$^\dag$ & \textbf{17.76}$^\dag$ & \textbf{31.83}$^\dag$ & 17.43 & \textbf{29.69}$^\dag$ & \textbf{23.69}$^\dag$ & \textbf{21.18}$^\dag$ & \textbf{22.43}$^\dag$   & 11.79                  & 5.10                  & 2.63                  & \textbf{7.90}$^\dag$   \\ 
\hline\hline
9                                                                  & \multirow{3}{*}{Conformer} & \xmark                    & \xmark                         & \multirow{3}{*}{\cmark}                                                & \multirow{3}{*}{\xmark}   & 48.71                 & 20.97                 & 36.93                 & 19.42 & 34.57                 & 33.08                 & 31.24                 & 32.15                   & 21.22                  & 6.63                  & 4.80                  & 13.72                  \\
10                                                                 &                            & \cmark                   & input layer                &                                                                     &                       & 47.86                 & 20.54                 & 36.57                 & 17.76 & \textbf{33.95}$^\dag$ & \textbf{31.88}$^\dag$ & 30.20                 & \textbf{31.04}$^\dag$   & 19.47                  & 6.73                  & 3.72                  & \textbf{12.53}$^\dag$  \\
11                                                                 &                            & \cmark                   & score A+AA                 &                                                                     &                       & \textbf{47.68}$^\dag$ & \textbf{20.38}$^\dag$ & 36.13                 & 17.87 & \textbf{33.75}$^\dag$ & \textbf{31.78}$^\dag$ & \textbf{29.93}$^\dag$ & \textbf{30.85}$^\dag$   & 19.43                  & 6.63                  & 3.64                  & \textbf{12.47}$^\dag$  \\
\hline\hline
12                                                                 & Sys. 8 $\rightarrow$ 10  & - & - & - & -         
& \textbf{38.96}$^\dag$  & \textbf{16.50}$^\dag$  & \textbf{29.77}$^\dag$ & \textbf{15.54}$^\dag$  & \textbf{27.58}$^\dag$  
& \textbf{22.87}$^\dag$  & \textbf{20.66}$^\dag$ & \textbf{21.76}$^\dag$  
& 10.41 & 3.88  & 3.02  & \textbf{7.04}$^\dag$   \\
\hline\hline
\end{tabular}
}
\vspace{-0.6cm}
\end{table*}

\vspace{-0.3cm}
\section{Experiments}
\vspace{-0.1cm}
\subsection{Experiments On Elderly Speech}
\vspace{-0.1cm}
\noindent
\textit{1) The DementiaBank Pitt corpus:} The DementiaBank Pitt \cite{becker1994natural} corpus contains about 33-hour audio recorded over interviews between the 292 elderly participants and clinical investigators. 
After silence stripping \cite{yu2018development}, the training set contains 15.7 hours of audio data (29682 utterances) while the development and evaluation sets contain 2.5 hours (5103 utterances) and 0.6 hours (928 utterances) of audio respectively. After a combination of speaker independent of elderly speech and dependent speed perturbation \cite{geng2020investigation} of non-aged investigators' speech based data augmentation, the duration of training data is increased to 58.9 hours (112830 utterances).
\par
\noindent \textit{2) The JCCOCC MoCA Corpus}: The Cantonese JCCOCC MoCA corpus \cite{xuspeaker} is based on speech recordings of cognitive impairment assessment interviews between 256 elderly participants and clinical investigators. After removal of excessive silence, the training set contains 32.1 hours of speech (95448 utterances) while the development and evaluation sets contain 3.5 hours (13675 utterances) and 3.4 hours (13414 utterances) of speech respectively. Data augmentation similar to those adopted on the DementiaBank Pitt corpus produces a 156.9 hours augmented training set (389409 utterances).
\par
\noindent \textit{3) Experiment Setup for the DementiaBank Pitt Corpus and the JCCOCC MoCA Corpus:} In our experiments, the 14-layer LF-MMI based TDNN-F models (17M) as shown in Figure 1 are implemented using the Kaldi toolkit \cite{povey2011kaldi} while Conformer based end-to-end systems\footnote{12 encoder+12 decoder layers, feed-forward layer dim=2048, attention heads=4, attention heads dim=256, CTC+AED cost.} (52M) are implemented using the Espnet toolkit \cite{watanabe2018espnet}. A 3-frame context window is used in both ASR and AASR hybrid LF-MMI trained TDNN systems. 40-dimensional Mel-scale filter banks (FBKs) are used as the input acoustic features. The 144-dimensional A2A generated UTI based articulatory features are produced by the approach described in section 3.2. For both the hybrid phonetic TDNN and E2E graphemic (character) Conformer systems, a word level 4-gram LM is used in recognition.
\par
\noindent \textit{4) Performance Analysis:} Domain adaptation is essential in transferring articulatory features from non-aged TaL speech to elderly corpora (Sys. 3 vs. 2). After incorporating the cross domain and cross lingual inverted UTI based articulatory features obtained on the DementiaBank Pitt or the JCCOCC MoCA audio data, the resulting AASR systems shown in Table 2 consistently outperform the comparable baseline ASR systems using acoustic features only by a significant margin in WER or CER (Sys. 3 vs. 1; Sys. 5 vs. 4; Sys. 7 vs. 6; Sys. 10 vs. 9) before and after data augmentation and LHUC speaker adaptation are applied. 
The performance improvements obtained by incorporating the inverted UTI features are retained after increasing the model complexity of the baseline TDNN ASR systems to be comparable to that of the AASR systems\footnote{ \#paramters of TDNN ASR system (Sys. 4) and larger one (Sys. 4.L) are 17M and 23M respectively, while that of TDNN AASR system (Sys. 5) is 23M, where the size of MLAN and inversion model is 6M.} (e.g. Sys. 4.L vs. 5).
In particular, significant WER reductions of 2.39\%-2.43\% are obtained on the elderly participants of the development and evaluation sets of the DementiaBank Pitt after data augmentation and LHUC adaptation of TDNNs (Sys. 7 vs. 6). Similar consistent improvement of 0.56\%-1.36\% on the JCCOCC MoCA corpus can be obtained (Sys. 7 vs. 6).
\par
Further investigation on the alternative forms of acoustic-articulatory modality fusion suggests that for both TDNN and Conformer based AASR systems, incorporating UTI based articulatory features via ``score A+AA'', a linear interpolation of system specific log-likelihood scores between ASR and AASR systems (Sys. 6 + 7, Sys. 9 + 10), produces further performance improvements over hidden layer or input layer feature fusion (Sys. 8 vs. 7, 11 vs. 10). After cross system 2-pass decoding\cite{cui2022two}, which uses the Conformer system (Sys. 10) to rescore the score fused AASR TDNN system (Sys. 8) outputs, overall WER and CER reductions of 4.75\% and 2.59\% over the baseline TDNN systems are obtained (Sys. 12 vs. 6) on the DementiaBank Pitt and the JCCOCC MoCA corpora respectively.
\par
Further ablation studies suggest that the best TDNN-F hidden layer to fuse the cross-domain and cross-lingual inverted UTI features vary across the English DementiaBank Pitt and the Cantonese JCCOCC MoCA data (7th vs. 3rd hidden layer). This may be attributed to the fact that the lower positioned TDNN-F layers model acoustic variability while the higher ones are more related to language specific phonetic targets. 

\vspace{-0.3cm}
\subsection{Experiments On Disordered Speech}
\vspace{-0.1cm}
\noindent \textit{1) the TORGO Corpus:} The TORGO dataset is a disordered speech corpus with 8 dysarthric and 7 control speakers. All 7 control speakers' data together with two thirds of the 8 dysarthric speakers' data are merged into the training set while the remaining dysarthric speech serves as the test data. After silence stripping and a combination of disordered speaker independent and control speaker dependent speed perturbation \cite{geng2020investigation} based data augmentation, the total training set contains 34.11 hours of data while the test set has 1.02 hours of speech.
\par

\noindent \textit{2) Experiment Setup for the TORGO Corpus:} The hybrid LF-MMI trained TDNN-F ASR system contains 7 hidden layers (9M), while the Conformer models\footnote{8 encoder+4 decoder layers, feed-forward layer dim=1024, attention heads=4, attention heads dim=256, CTC+AED cost.} (18M) are trained to directly model grapheme (character) sequence outputs. A 3-frame context window is used in both ASR and AASR hybrid TDNN-F systems. 40-dimensional Mel-scale filter banks (FBKs) and 144-dimensional UTI based features are used as the input acoustic and articulatory features respectively. A 3-gram LM trained by all the TORGO transcripts is used in decoding. 
\par

\noindent \textit{3) Performance Analysis: } Several trends similar to those found on the two elderly speech tasks can be observed. Compared with ASR systems, a significant overall WER reduction of 1.02\% absolute (11.20\% relative) can be obtained after data augmentation and speaker adaptation (Sys. 7 vs. 6). The AASR Conformer outperforms the comparable ASR baseline significantly by 1.19\% absolute (8.67\% relative) (Sys.10 vs. 9). Further with score fusion and cross system multi-pass decoding, the lowest overall WER of 7.04\% is obtained (Sys. 12).

\vspace{-0.3cm}
\subsection{T-SNE Visualization And RMSE}
\vspace{-0.1cm}
Further analysis over the phonetic discrimination brought by UTI articulatory features via t-distributed stochastic neighbour embedding (t-SNE) \cite{van2008visualizing} visualization is shown in the Figure 2 across four datasets. 
As expected, the original UTI articulatory features produce clear phonetic discrimination between example phonemes /\i/ and /\textopeno/ on the TaL test set. Similar strong phonetic discrimination obtained using the cross-domain and cross-lingual generated UTI features can also be found in Figure 2 (b), (c) and (d) for the same pair of phonemes on the English DementiaBank Pitt, the Cantonese JCCOCC MoCA\footnote{UTI features are computed from tonal Cantonese phonemes with identical base phoneme but different tones are averaged.} and the English TORGO test data. In addition, comparable RMSE scores (1.027, 1.030, 1.025 and 1.028) were obtained when performing in-domain and cross-domain UTI inversion for the TaL, DementiaBank Pitt, JCCOCC MoCA and TORGO corpora respectively. This demonstrates the robustness of our cross-domain and cross-lingual UTI inversion approach. 
In Figure 3, a further t-SNE plot of the generated articulatory features computed over phonemes /b/ and /d/ on the TORGO data with and without MLAN illustrates the importance of cross-domain and cross-lingual adaptation in UTI inversion.

\vspace{-0.2cm}

\vspace{-0.1cm}
\section{Conclusion}
\vspace{-0.1cm}
This paper presents a cross-domain and cross-lingual A2A inversion approach that utilizes the parallel audio and ultrasound tongue imaging (UTI) data of the TaL corpus in A2A model training before being cross-domain and cross-lingual adapted to three datasets across two languages: English DementiaBank Pitt and Cantonese JCCOCC MoCA elderly speech corpora; and English TORGO dysarthric speech data to produce UTI based articulatory features. On three tasks, incorporating the A2A generated articulatory features consistently and significantly outperforms the baseline hybrid TDNN and end-to-end Conformer ASR systems constructed using acoustic features only. The proposed cross-domain and cross-lingual A2A inversion method allows more practical and wider use of UTI based articulatory features in elderly and disordered ASR systems.

\vspace{-0.5cm}
\begin{figure}[htbp]
  \centering
  \subfigure[the TaL corpus]{
  \begin{minipage}[t]{0.46\linewidth}
    \centering
    \includegraphics[width=\linewidth]{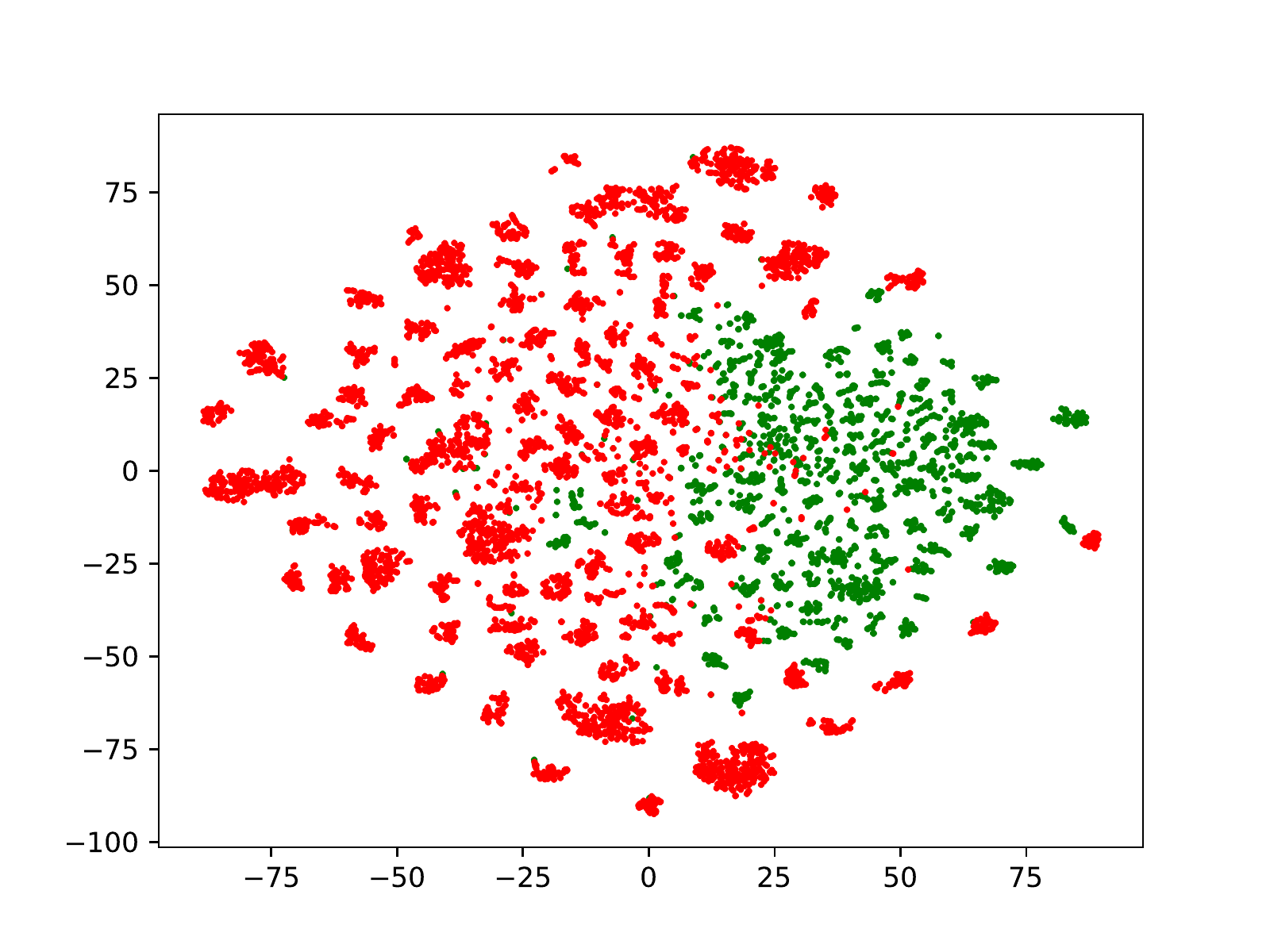}
    \vspace{-0.6cm}
    \label{fig:side:a}
  \end{minipage}%
  }
  \subfigure[the DementiaBank Pitt corpus]{
  \begin{minipage}[t]{0.46\linewidth}
    \centering
    \includegraphics[width=\linewidth]{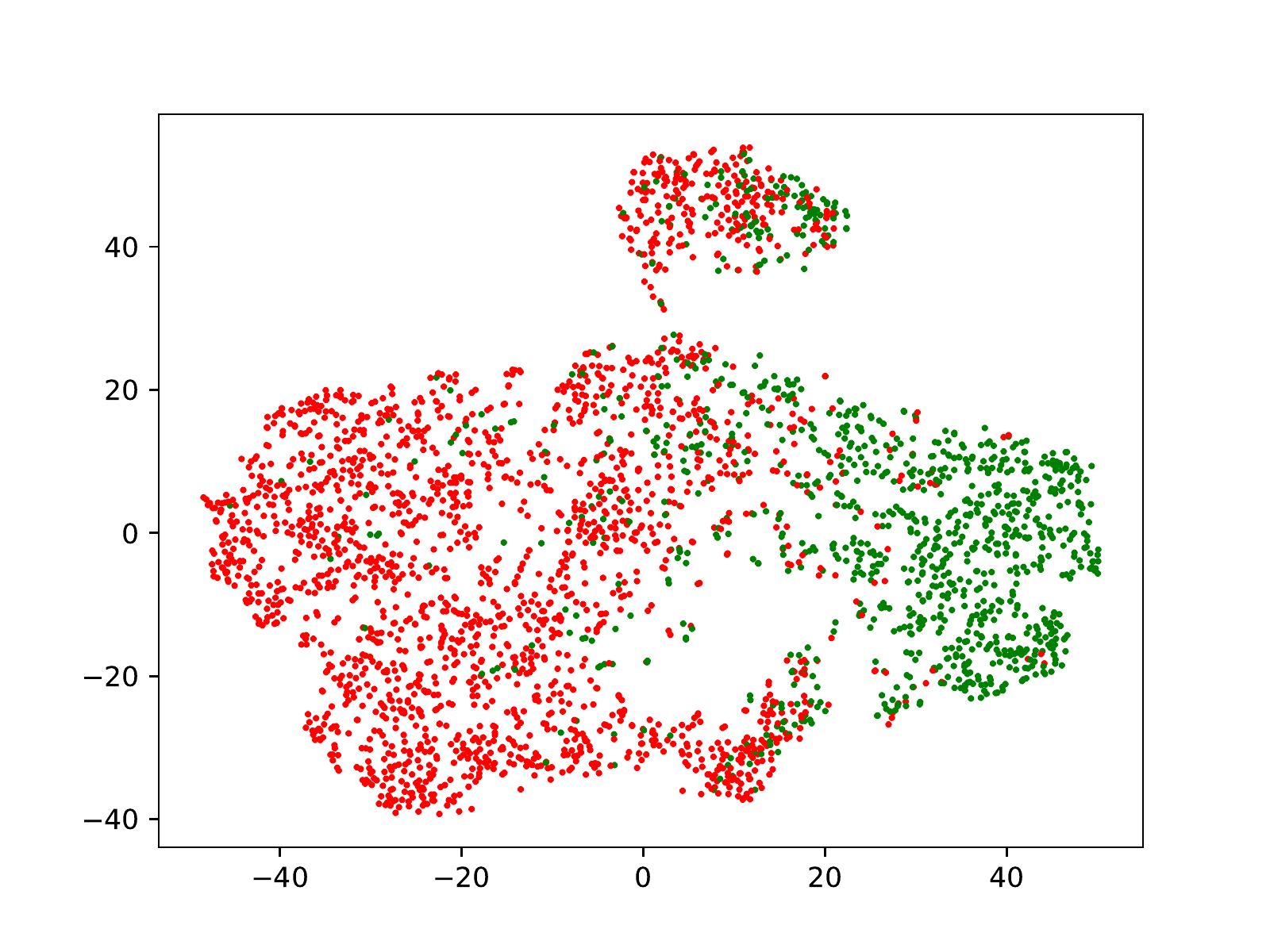}
    \vspace{-0.6cm}
    \label{fig:side:b}
  \end{minipage}
  }
  \subfigure[the JCCOCC MoCA corpus]{
  \begin{minipage}[t]{0.46\linewidth}
    \centering
    \includegraphics[width=\linewidth]{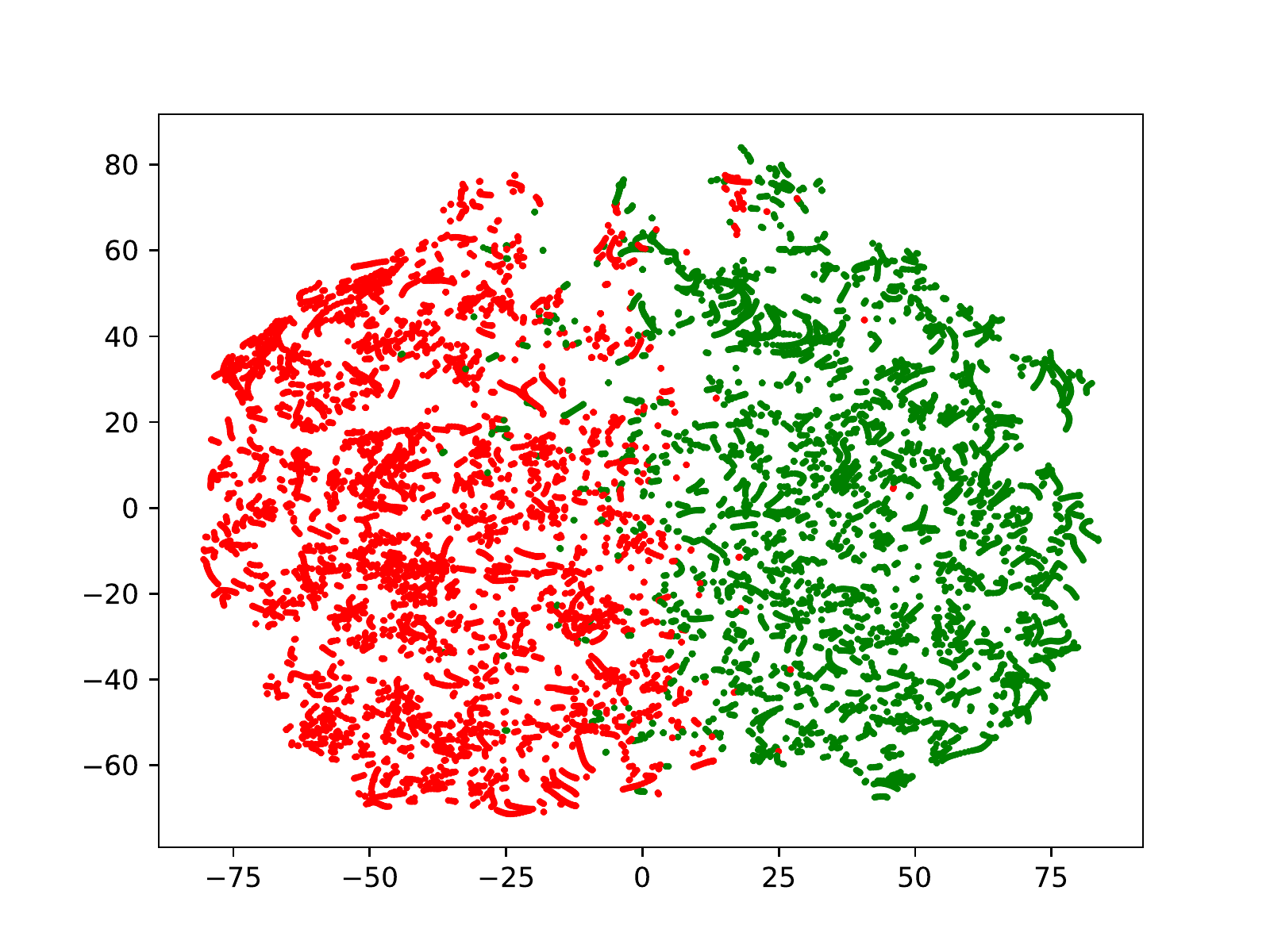}
    \vspace{-0.6cm}
    \label{fig:side:c}
  \end{minipage}
  }
  \subfigure[the TORGO corpus]{
  \begin{minipage}[t]{0.46\linewidth}
    \centering
    \includegraphics[width=\linewidth]{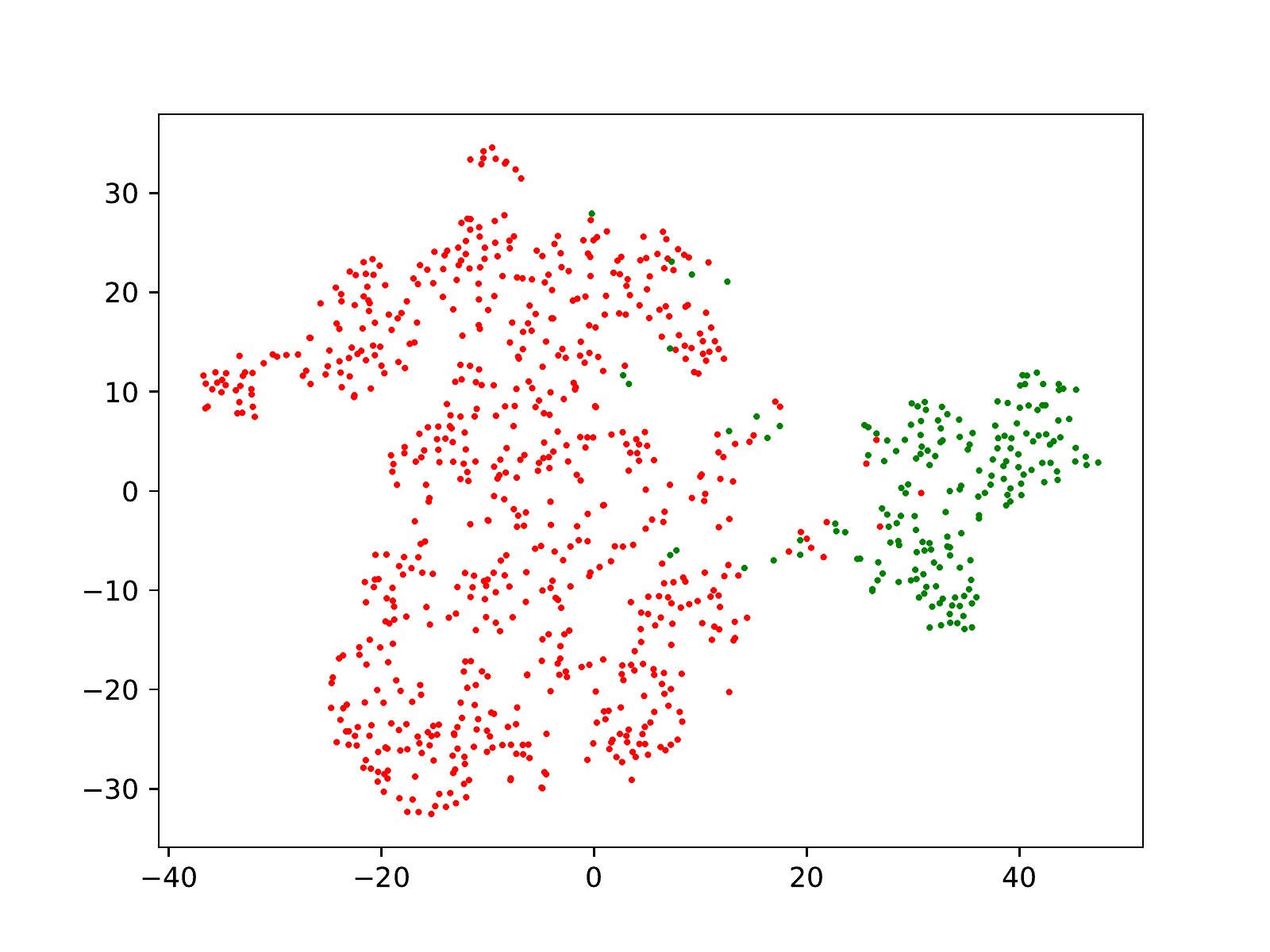}
    \vspace{-0.6cm}
    \label{fig:side:d}
  \end{minipage}
  }
  \vspace{-0.4cm}
  \caption{T-SNE plot of original (a) and cross-domain/lingual generated ((b) to (d)) UTI articulatory features computed over phonemes /\i/ (in red) and /\textopeno/ (in green) on the TaL, DementiaBank Pitt, JCCOCC MoCA and TORGO corpora.}
  \vspace{-0.2cm}
\end{figure}
\vspace{-0.8cm}
\begin{figure}[H]
  \centering
  \subfigure[without MLAN]{
  \begin{minipage}[t]{0.48\linewidth}
    \centering
    \includegraphics[width=\linewidth]{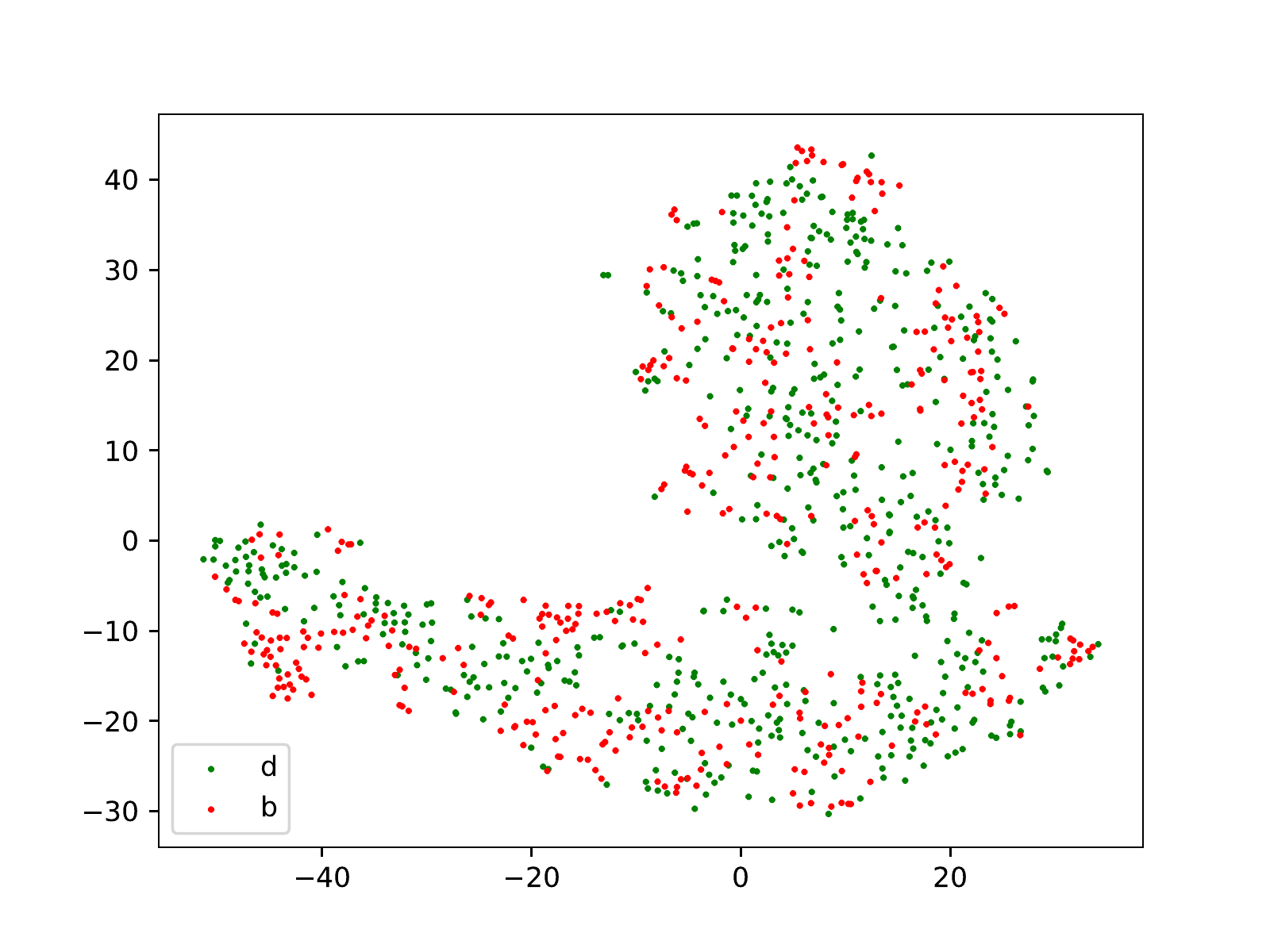}
    \vspace{-0.6cm}
    \label{fig:side:a}
  \end{minipage}%
  }
  \subfigure[with MLAN]{
  \begin{minipage}[t]{0.48\linewidth}
    \centering
    \includegraphics[width=\linewidth]{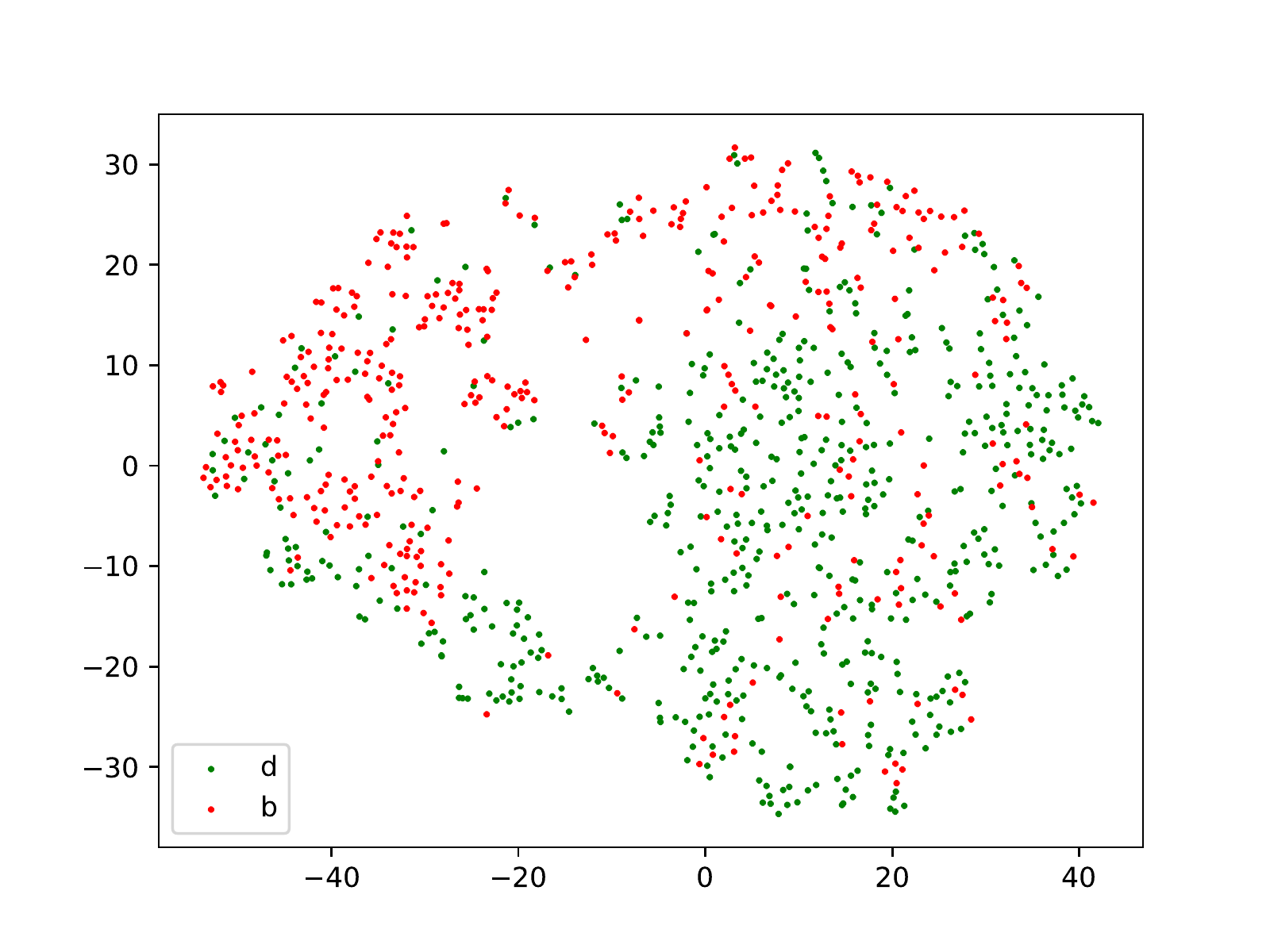}
    \vspace{-0.6cm}
    \label{fig:side:b}
  \end{minipage}
  }
  
  \vspace{-0.4cm}
  \caption{T-SNE plot of generated UTI articulatory features computed over phonemes /b/ (in red) and /d/ (in green) on TORGO data w/o cross-domain and cross-lingual adaptation.}
  \vspace{-0.2cm}
\end{figure}
\vspace{-0.7cm}

\section{Acknowledgement}
\vspace{-0.1cm}
This research is supported by Hong Kong RGC GRF grant No. 14200220, 14200021, TRS T45-407/19N, Innovation \& Technology Fund grant No. ITS/218/21, National Natural Science Foundation of China (NSFC) Grant 62106255 and Youth Innovation Promotion Association CAS Grant 2023119.

\bibliographystyle{IEEEtran}
\bibliography{mybib}
\end{document}